\documentclass[aps,twocolumn,preprintnumbers,floatfix,pra,10pt]{revtex4-2}
\usepackage{bbm}
\usepackage{bm}
\usepackage{amsmath}
\usepackage{amssymb}
\usepackage{empheq}
\usepackage{graphicx}
\usepackage{mathrsfs}
\usepackage{amsfonts}
\usepackage{amsthm}
\usepackage{color}
\usepackage{multirow}
\usepackage{bigints}
\usepackage{txfonts}
\usepackage{float}
\usepackage{hyperref}
\hypersetup{
     unicode=false,          
     pdftoolbar=true,        
     pdfmenubar=true,        
     pdffitwindow=false,     
     pdfstartview={FitH},    
     pdftitle={My title},    
     pdfauthor={Author},     
     pdfsubject={Subject},   
     pdfcreator={Creator},   
     pdfproducer={Producer}, 
     pdfkeywords={keyword1} {key2} {key3}, 
     pdfnewwindow=true,      
     colorlinks=false,       
     linkcolor=red,          
     citecolor=green,        
     filecolor=magenta,      
     urlcolor=cyan           
}

\makeatletter \tolerance = 10000 \tolerance = 10000
\usepackage{color}

\makeatother
\begin{document}

\title{Chiral Hinge Transports in Disordered Non-Hermitian Second-Order Topological Insulators}

\author{C. Wang}
\email[Corresponding author: ]{physcwang@tju.edu.cn}
\affiliation{Center for Joint Quantum Studies and Department of Physics, School of Science, Tianjin University, Tianjin 300350, China}
\author{X. R. Wang}
\email[Corresponding author: ]{phxwan@ust.hk}
\affiliation{Physics Department, The Hong Kong University of Science and Technology (HKUST), Clear Water Bay, Kowloon, Hong Kong}
\affiliation{HKUST Shenzhen Research Institute, Shenzhen 518057, China}

\date{\today}

\begin{abstract}

The generalized bulk-boundary correspondence predicts the existence of the chiral hinge states in three-dimensional second-order topological insulators (3DSOTIs), resulting in a quantized Hall effect in three dimensions. Chiral hinge states in Hermitian 3DSOTIs are characterized by the quantized transmission coefficients with zero fluctuations, even in the presence of disorders. Here, we show that chiral hinge transport in disordered non-Hermitian systems deviates from the paradigm of the Hermitian case. Our numerical calculations prove the robustness of hinge states of disordered non-Hermitian 3DSOTIs. The mean transmission coefficients may or may not equal the number of chiral hinge channels, depending on the Hermiticity of chiral hinge states, while the fluctuations of transmission coefficients are always non-zero. Such fluctuations are not due to the broken chirality of hinge states but the incoherent scatterings of non-Hermitian potentials. The physics revealed here should also be true for one-dimensional chiral channels in topological materials that support chiral boundary states, such as Chern insulators, three-dimensional anomalous Hall insulators, and Weyl semimetals.
 
\end{abstract}

\maketitle

\section{Introduction}
\label{sec1}

Searching various topological phases has attracted tremendous attention because of their fundamental interests and exotic properties 
for potential applications~\cite{haldane_prl_1988,kane_prl_2005,bernevig_science_2006,konig_science_2007,change_science_2013,
xu_science_2015,lu_science_2015,fu_prl_2008,qi_prl_2009,grover_science_2014}. One such feature is the emergence of 
one-dimensional (1D) chiral channels along the edge of two-dimensional topological insulating systems at the Fermi level~\cite{hasan_rmp_2010,bansil_rmp_2016,chiu_rmp_2016}. Chiral edge states result in quantized conductances for the well-known 
quantum anomalous Hall effect paradigm~\cite{change_science_2013}. Recently, the study has been expanded to the higher-order 
topological insulators that follow the generalized bulk-boundary correspondence~
\cite{benalcazar_science_2017,langbehn_prl_2017,song_prl_2017,ezawa_prl_2018,liu_prl_2019,kudo_prl_2019,li_npj_2019,chen_prl_2020,
saaghorashi_prb_2021_1,saaghorashi_prb_2021_2}. One widely investigated example is the strong three-dimensional second-order 
topological insulator (3DSOTI) that supports 1D chiral hinge states localized at the intersection of two surfaces~
\cite{zhang_prl_2013,benalcazar_prb_2017,schindler_sciadv_2018}. 
\par

The robustness of these topological phases against disorders is universal~\cite{prodan_prl_2010,liu_prl_2016,liu_nm_2020}. 
Concerning chiral hinge states in 3DSOTIs, on the one hand, their robustness has been recognized in the study of Hermitian 
systems~\cite{wang_prresearch_2020,zdsong_prl_2021,jhwang_prl_2021,plzhao_prl_2021,wang_prb_2021}. Consequently, 
disordered Hermitian 3DSOTIs are characterized by the quantized transmission coefficients (equal to the number of chiral 
channels) and the vanishingly-small fluctuations due to the absence of backward scattering. On the other hand, for non-Hermitian 
systems, the widely accepted paradigm asserts no quantized transport for chiral edge states in Chern (first-order topological) 
insulators~\cite{philip_prb_2018,chen_prb_2018,groenendijk_prr_2021}. The effective Hamiltonians of such chiral edge 
states are non-Hermitian, and the corresponding eigenenergies are complex numbers. Recently, we have shown that the chiral 
hinge states of a non-Hermitian 3DSOTI can be Hermitian under certain conditions~\cite{cwang_prb_2022}. So far, the fate 
and the transport of chiral hinge states in non-Hermitian 3DSOTIs under disorders are still open questions, especially for the 
Hermitian chiral hinge states.
\par

Here, we numerically investigate the nature of chiral hinge states of disordered 3DSOTIs subject to constant non-Hermitian 
and random Hermitian on-site potentials. The main findings are as follows: Both Hermitian and non-Hermitian chiral 
hinge states can persist in the presence of finite disorders. As the disorder strength increases, the non-Hermitian 3DSOTI 
undergoes first a transition to the non-Hermitian three-dimensional first-order topological insulator (3DFOTIs) featured by 
the surface states and then follow by another transition to the topologically-trivial metal. The mean transmission coefficients 
of Hermitian chiral hinge states of non-Hermitian 3DSOTIs are still equal to the number of chiral hinge channels, similar to 
the chiral hinge states of the Hermitian 3DSOTIs~\cite{wang_prresearch_2020}, but the transmission coefficients of of the 
non-Hermitian chiral hinge states can be any number. In both cases, the fluctuations of transmission coefficients are 
finite and increase with the disorders and the system sizes. Furthermore, the fluctuations of transmission coefficients 
are not due to the broken chirality of hinge states but from the incoherent scattering by non-Hermitian potentials. The 
distributions of transmission coefficients of chiral hinge states follow universal forms that depend on the Hermiticity of 
chiral hinge states. Our findings highlight the importance of the incoherent processes in non-Hermitian topological 
insulators, which are often ignored in the literature~\cite{philip_prb_2018,chen_prb_2018,groenendijk_prr_2021}, and 
should apply to other non-Hermitian topological insulators that support chiral boundary states.  
\par

The paper is organized as follows: A tight-binding model for the non-Hermitian 3DSOTI is introduced in Sec.~\ref{sec2}.  Non-Hermiticities effect on chiral hinge states in the clean limit is discussed. The study of disorder-driven the quantum phase transitions of non-Hermitian 3DSOTIs is presented in Sec.~\ref{sec3}. Section~\ref{sec4} gives a comprehensive investigation of quantum transports through chiral hinge states in non-Hermitian 3DSOTIs followed by the discussion in Sec.~\ref{sec5}. The conclusion is given in Sec.~\ref{sec6}.
\par 

\section{Tight-binding models and clean systems}
\label{sec2}

We consider following tight-binding model on the cubic lattices
\begin{equation}
\begin{gathered}
H=\sum_{\bm{i}}c^\dagger_{\bm{i}}\left[ v_{\bm{i}}\Gamma^0+M\Gamma^4+B\Gamma^{32}+i\sum_{\mu=0,1,2,3,4,5}\gamma_\mu\Gamma^\mu \right]c_{\bm{i}}\\
+\sum_{\bm{i}}\sum_{\nu=1,2,3}\left[\left(\dfrac{it}{2}c^\dagger_{\bm{i}+\hat{x}_\nu}\Gamma^{\nu}-\dfrac{t}{2}c^\dagger_{\bm{i}+\hat{x}_\nu}\Gamma^{4}\right)c_{\bm{i}}+h.c. \right],
\end{gathered}\label{eq2_1}
\end{equation}
which belongs to the class A of the Altland-Zirnbauer (AZ) classifications for non-Hermitian systems~\cite{kawabata_prx_2019}. Here, $c^\dagger_{\bm{i}}$ and $c_{\bm{i}}$ are the single-particle creation and annihilation operators at site $\bm{i}=(n_1,n_2,n_3)$. Particles governed by Eq.~\eqref{eq2_1} can be either fermions like electrons or bosons like magnons. $\hat{x}_{\nu=1,2,3}$ are unit vectors along three Cartesian axes. $\Gamma^{\mu=1,2,3,4,5}=(s_0\otimes\sigma_3,s_1\otimes\sigma_1,s_3\otimes\sigma_1,s_2\otimes\sigma_1,s_0\otimes\sigma_2)$ are four-by-four nonunique gamma matrices that satisfy $\{\Gamma^{\mu},\Gamma^{\nu}\}=2\delta_{\mu,\nu}\Gamma^0$ and $\Gamma^{\mu\nu}=[\Gamma^\mu,\Gamma^\nu]/(2i)$, where $\Gamma^0$ and $s_0,\sigma_0$ are four-by-four and two-by-two unit matrices, respectively, and $s_\mu,\sigma_\mu$ are the Pauli matrices acting on some subspaces (e.g., spins or orbitals).  $t$ is the hopping energy and is used as the energy unit ($t=1$) here. $M$ and $B$ are the masses for controlling the band inversions of bulk states and surface states, respectively. The non-Hermiticity is encoded in the on-site potential $\sum_{\bm{i}}c^\dagger_{\bm{i}}(i\sum_{\mu=0,1,2,3,4,5}\gamma_\mu\Gamma^\mu)c_{\bm{i}}$ with $\gamma_{0,1,2,3,4,5}$ measure the strengths of non-Hermitian potentials. $v_{\bm{i}}$ is a white noise that distributes uniformly in the range of $[-W/2,W/2]$.
\par

In the absence of the disorder, our model can be blocked diagonalized in the $\bm{k}$ space due to the Bloch's theorem, $H=\sum_{\bm{k}}a^\dagger_{\bm{k}}h(\bm{k})a_{\bm{k}}$ with 
\begin{equation}
\begin{gathered}
h(\bm{k})=t\sum_{\mu=1,2,3}\sin k_\mu\Gamma^\mu+\left(M-t\sum_{\mu=1,2,3}\cos k_\mu\right)\Gamma^4+B\Gamma^{32}\\
+i\sum_{\mu=0,1,2,3,4,5}\gamma_\mu\Gamma^\mu.
\end{gathered}\label{eq2_2}
\end{equation}
Without the $\gamma$-terms, Hamiltonian~\eqref{eq2_2} describes a Hermitian reflection-symmetric 3DSOTI with the reflection plane of $x=0$ for $0<B<1$  and $1+B<M<3-B$ \cite{wang_prresearch_2020}. Chiral hinge states appear at the boundaries when two surfaces meet at the reflection plane. Namely, if we apply Eq.~\eqref{eq2_2} to a cubic lattice of size $L_\parallel/\sqrt{2}\times L_\parallel/\sqrt{2}\times L_\perp$ with open boundary conditions (OBCs) on surfaces perpendicular to $(110),(1\bar{1}0),(001)$, the in-gap chiral hinge modes are found at the two hinges $x=0,y=\pm L_\parallel$.
\par 

\begin{figure}[htbp]
\centering
\includegraphics[width=0.45\textwidth]{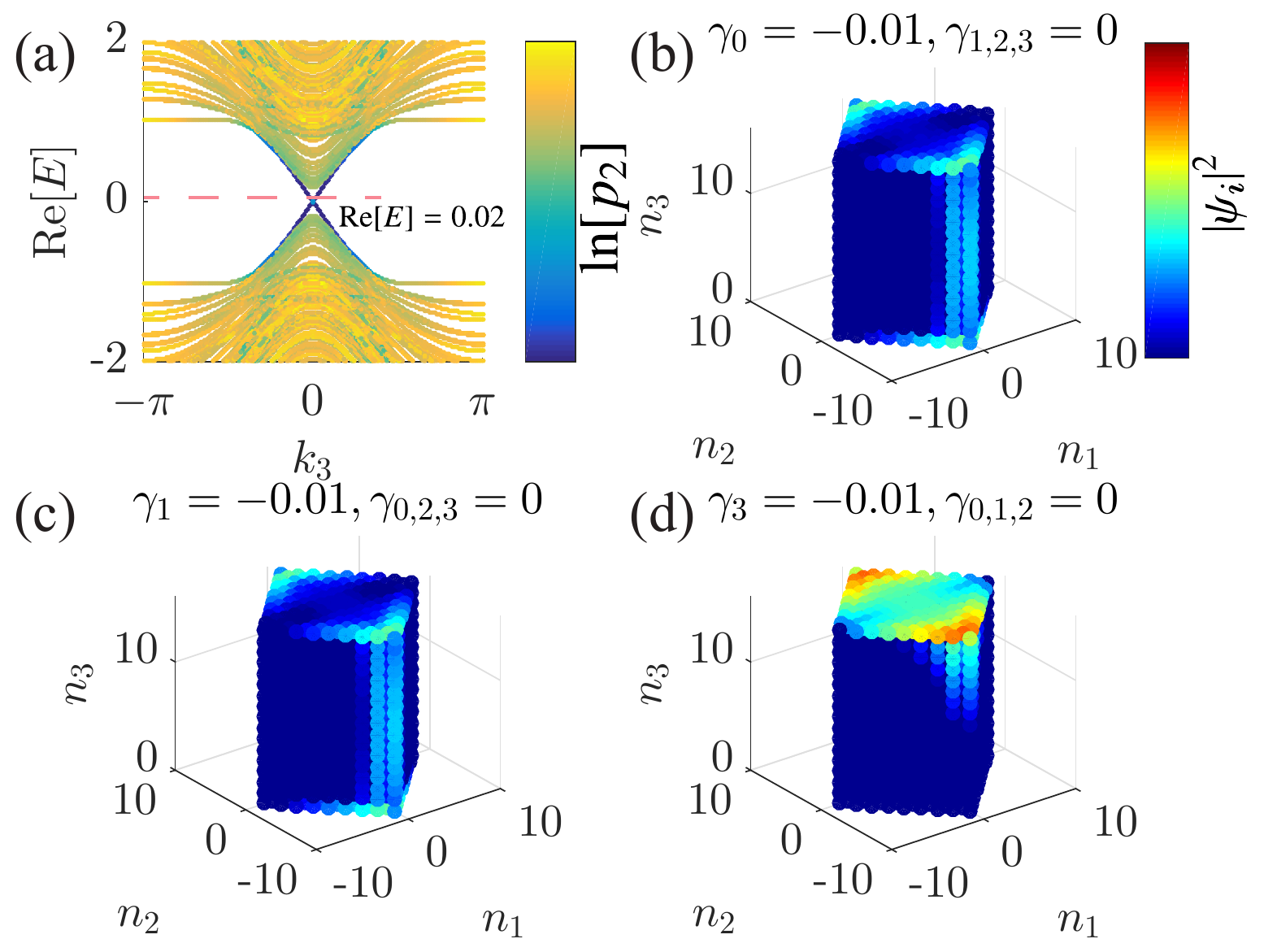}
\caption{(a) $\text{Re}[E](k_3)$ of Eq.~\eqref{eq2_1} for $M=2$, $B=0.2$, $\gamma_{0,1,2,3,4,5}=0$ and $L_\parallel=24$. OBCs are applied in surfaces perpendicular to $(110)$ and $(1\bar{1}0)$, PBC is along the $z-$direction. Colors map $\ln[p_2]$. Dash line locates the state with $\text{Re}[E]=0.02$. (b,c,d) Spatial distributions of wave functions $|\psi_{\bm{i}}|^2=\sum^4_{p=1}|\psi_{\bm{i},p}|^2$ of Eq.~\eqref{eq2_1} for (b) $\gamma_0=-0.01,\gamma_{1,2,3}=0$, (c) $\gamma_1=-0.01,\gamma_{0,2,3}=0$, and (d) $\gamma_3=-0.01,\gamma_{0,1,2}=0$. $\psi_{\bm{i}}$ are the corresponding normalized right eigenfunctions of $E=0+i\gamma_0$. $M=2$, $B=0.2$, and $L_\parallel=L_\perp=16$. Here, we set OBCs in all surfaces.}
\label{fig1}
\end{figure}

The chiral hinge states are visible in Fig.~\ref{fig1}(a) which displays the energy spectrum of quasiparticles $\text{Re}[E](k_3)$ of Eq.~\eqref{eq2_1} of $M=2$, $B=0.2$, and $\gamma_{0,1,2,3,4,5}=0$ for a rectangle bar with OBCs on the surfaces perpendicular to $(110)$ and $(1\bar{1}0)$ and periodic boundary condition (PBC) in the $z-$direction. Colors in Fig.~\ref{fig1}(a) encode the natural logarithm of the participation ratios $\ln[p_2]$. The participation ratio is defined as $p_2(E)=1/(\sum_{\bm{i}}\sum^{4}_{p=1}|\psi_{\bm{i},p}(E)|^4)$ with $\psi_{\bm{i},p}(E)$ being the wave function of a right eigenstate of $H$ of a site $\bm{i}$. The participation ratio measures the numbers of occupied sites of a given state of the eigenvalue $E$~\cite{cwang_prl_2015}. $\ln[p_2]$ for the in-gap hinge states is small (in comparison with the number of lattice sites), see Fig.~\ref{fig1}(a). 
\par

Non-Hermiticities are introduced through the $\gamma$-terms in Eq.~\eqref{eq2_2}. $i\gamma_0 \Gamma^0$ of $\gamma_0<0$ ($\gamma_0>0$) describes a non-local quantum incoherent loss (gain) of the chiral hinge states. $i\gamma_j\Gamma^j$ of $j=1,2,3$, acting on the eigenstates of the Hermitian part, adds a pre-factor $\exp[\gamma_j x_j/t]$ to the Bloch wave function of the 3DSOTI and leads to the non-Hermitian skin effect~\cite{yao_prl_2018}. The remaining two terms, $i\gamma_4\Gamma^4$ and $i\gamma_5\Gamma^5$, make the Dirac masses of bulk and surface states complex such that the phase boundaries between the 3DSOTI and the second-order Weyl semimetals are tuned by the non-Hermitian potentials~\cite{wang_prresearch_2020}. In what follows, we ignore the two terms $i\gamma_4\Gamma^4$ and $i\gamma_5\Gamma^5$.
\par

Through the same approach used in Ref.~\cite{cwang_prb_2022}, we derive the effective low-energy continuum Hamiltonian of chiral hinge states 
\begin{equation}
\begin{gathered}
h_{\text{hinge}}(p_3)=i\gamma_0\sigma_0+t(p_3+i\gamma_3/t)\sigma_3
\end{gathered}\label{eq2_3}
\end{equation}
in the basis of the zero-energy hinge states localized at the two hinges $x=0,y=\pm L_\parallel/2$. The Hermitian part of $tp_3\sigma_3$ gives two chiral hinge channels propagating in opposite directions. The $\gamma_0$-term is a non-local gain or loss if $\gamma_0$ is positive or negative, respectively. While eigenstates of $h_{\text{hinge}}(\gamma_3=0)$ are extended on the two hinges, $h_{\text{hinge}}(\gamma_3\neq 0)$'s eigenstates are exponentially localized at the end of the hinges, i.e., the corners. As we illustrated in Ref.~\cite{cwang_prb_2022}, the hinge states of a non-Hermitian 3DSOTI will be Hermitian if neither the gain/loss nor the non-Hermitian skin effect is presented, say $\gamma_0=\gamma_3=0$ and $\gamma_{1,2}\neq 0$. 
\par

The above picture can be seen by the spatial distributions of wave functions of a right eigenstate of Eq.~\eqref{eq2_1} on a lattice of size $L_\parallel=L_\perp=16$ for three cases: (i) $\gamma_0=-0.01,\gamma_{1,2,3}=0$, and $E=0-0.01i$; (ii) $\gamma_1=-0.01,\gamma_{0,2,3}=0$, and $E=0$; (iii) $\gamma_3=-0.01,\gamma_{0,1,2}=0$, and $E=0$. The distributions are shown in Figs.~\ref{fig1}(b,c,d). While the wave functions of hinge states spread over the whole hinges for cases (i) and (ii), it is localized at the two corners $x=0,y=\pm L_\parallel,z=L_\perp$ for the case (iii) where $\gamma_3=-0.01$ and $\gamma_{0,1,2}=0$. Such corner states are due to the non-Hermitian skin effect on hinge states and will localized at $x=0,y=\pm L_\parallel,z=-L_\perp$ if we choose $\gamma_3=0.01$ and $\gamma_{0,1,2}=0$. Since the focus of this work is on the chiral hinge states, we set $\gamma_3=0$ and only consider cases (i) and (ii) in what follows.
\par

\section{Phase diagram of non-Hermitian disordered 3DSOTI}
\label{sec3}

Now, we concentrate on the fate of hinge states of non-Hermitian 3DSOTIs in the presence of finite disorders. To this end, we turn on the disorder potentials $W>0$ and consider a specific energy $E=0.02+i\gamma_0$ and $M=2,B=0.2$ in Eq.~\eqref{eq2_1}. As shown in Fig.~\ref{fig1}(a), the $\text{Re}[E]=0.02$ state is a chiral hinge state in the clean limit. In the presence of disorders, if a state of  a complex eigenvalue $E$ is still the hinge state, its wave function should be highly concentrated on the hinges. We thus use the following quantity to identify a hinge state: 
\begin{equation}
\begin{gathered}
\eta_{W,L}(E)=\sum_{\bm{j}\in\text{hinge}}\sum^4_{p=1}|\psi_{\bm{j},p}(E)|^2
\end{gathered}\label{eq3_1}
\end{equation}
with the first summation over all lattice sites on the two hinges ($x=0,y=\pm L_\parallel/2$) and $|\psi_{\bm{j},p}(E)|$ being the wave function amplitude of a right eigenstate $|\psi(E)\rangle$ of the eigenvalue $E$ on site $\bm{j}$. Here, we choose the normalization condition $\langle\psi(E)|\psi(E)\rangle=1$. The identification of a hinge state is guided by the following features~\cite{wang_prresearch_2020,wang_prb_2021}: For hinge states, $\eta_{W,L}$ approaches a finite value for $L_\parallel=L_\perp=L\gg 1$, while $\eta_{W,L}$ decreases with $L$ for surface and bulk states. Numerically, we compute $\eta_{W,L}$ through the exact diagonalization by using the Scipy library~\cite{scipy} and the KWANT package~\cite{kwant}.     
\par

\begin{figure}[htbp]
\centering
\includegraphics[width=0.45\textwidth]{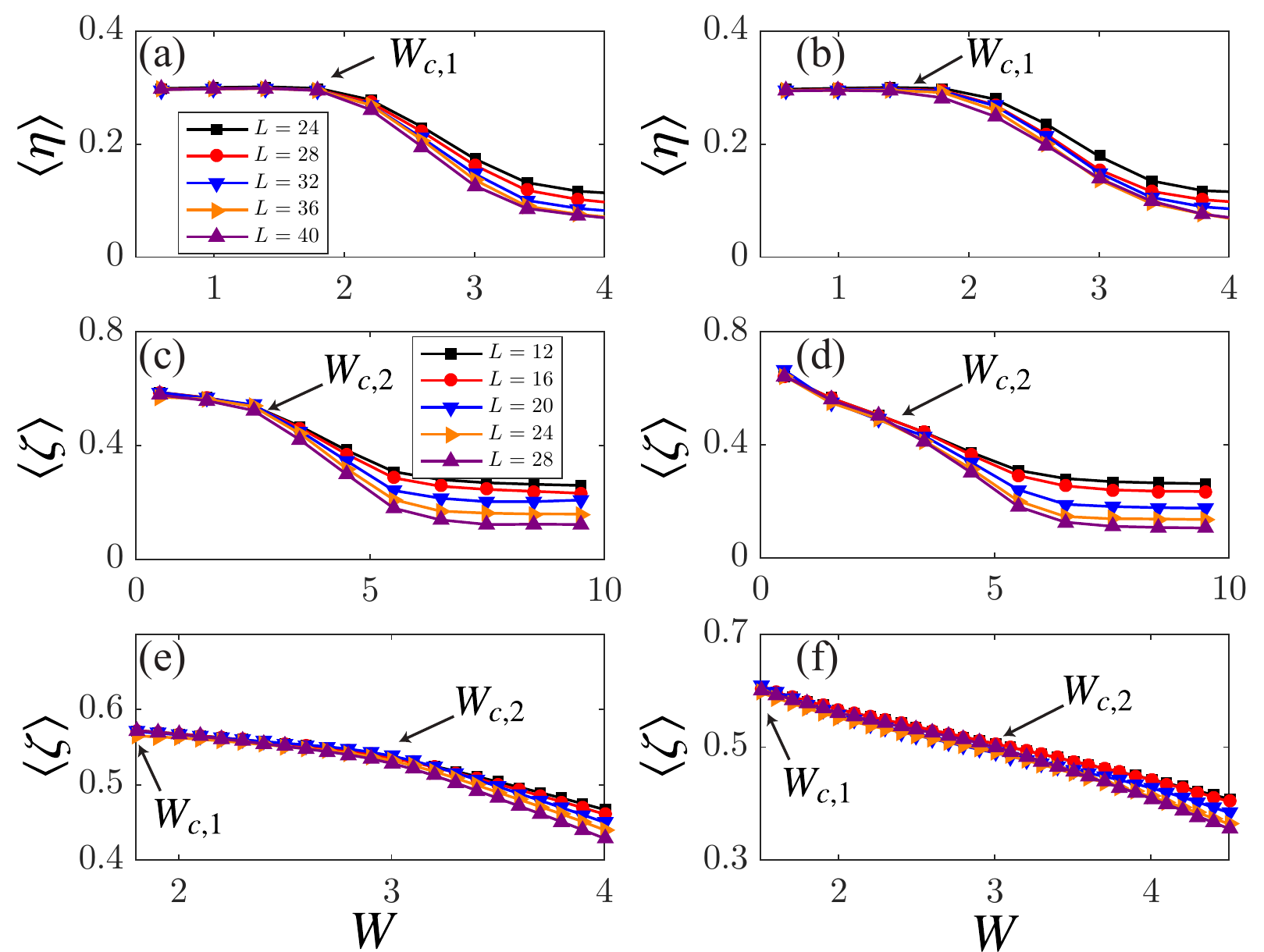}
\caption{(a,b) $\langle\eta \rangle$ as a function $W$ for $L=24,28,32,36,40$ with (a) $\gamma_0=-0.01,\gamma_{1,2,3}=0$, and $E=0.02-i0.01$; (b) $\gamma_1=-0.01,\gamma_{0,1,2}=0$, and $E=0.02$. The arrows locate the positions of $W_{c,1}$ schematically. (c,d) $\langle\zeta \rangle$ as a function $W$ for $L=12,16,20,24,28$ with (c) $\gamma_0=-0.01,\gamma_{1,2,3}=0$ and (d) $\gamma_1=-0.01,\gamma_{0,2,3}=0$. The arrows point to $W_{c,2}$.
(e,f) Zoom in on $\langle \zeta\rangle$ in the regime of the 3DFOTI phase ($W\in [W_{c,1},W_{c,2}]$) for (e) $\gamma_0=-0.01,\gamma_{1,2,3}=0$ and (f)  $\gamma_1=-0.01,\gamma_{0,2,3}=0$.}
\label{fig2}
\end{figure}

Figures~\ref{fig2}(a,b) display $\langle\eta_{W,L}\rangle$ of $E=0.02+i\gamma_0$ for various $W$ and $L$ of cases (i) $\gamma_0=-0.01,\gamma_{1,2,3}=0$; (ii) $\gamma_1=-0.01,\gamma_{0,2,3}=0$; where the hinge states are non-Hermitian (dissipative) and Hermitian for $W=0$, see Eq.~\eqref{eq2_3}. The symbol $\langle\cdot\rangle$ stands for the ensemble average. For both non-Hermitian and Hermitian hinge states, curves $\langle \eta_{W,L} \rangle$ of different $L$ merge at $W=W_{c,1}$ for $W\leq W_{c,1}$ and increases with the decreases of $L$ for $W>W_{c,1}$. For dissipative chiral hinge states, $W_{c,1}\simeq 1.8$, and for Hermitian chiral hinge states $W_{c,1}\simeq 1.4$. Hence, in each case, chiral hinge states can survive at finite disorders, which commensurate with the observation in Hermitian systems~\cite{wang_prresearch_2020,wang_prb_2021}. 
\par

The 3DSOTI-to-3DFOTI transitions can be understood as follows. In the absence of disorders, there are three types of states in a non-Hermitian 3DSOTI: topological hinge states, topological surface states, and trivial bulk states. In Ref.~\cite{cwang_prb_2022}, we have shown that topological hinge states in the complex energy gap of topological surface states and topological surface states in the bulk energy gap are due to the band inversion of surface and bulk states, respectively. Disorders tend to close the energy gap of surface states first and then bulk states. Therefore, the disorder-induced quantum phase transitions occur first from non-Hermitian 3DSOTIs to 3DFOTIs, then from 3DSOTIs to 3D non-hermitian metals. The numerical verification of the assertion for case (i) is presented in Appendix~\ref{sec_transitions}.

Naturally, we expect that a 3DFOTI-to-3D-metal transition occurs at a stronger disorder $W_{c,2}>W_{c,1}$ where the gap of bulk states closes. Since the 3DFOTIs are featured as bulk-gap states localized on the surfaces, the following quantity 
\begin{equation}
\begin{gathered}
\zeta_{W,L}(E)=\sum_{\bm{j}\in\text{surface}}\sum^4_{p=1}|\psi_{\bm{j},p}(E)|^2,
\end{gathered}\label{eq3_2}
\end{equation}
which describes the occupation probability, or inverse participation ration, of a state on the surfaces or hinges of a sample, should be independent of $L$ for surface/hinge states and decrease with $L$ for the bulk states~\cite{wang_prresearch_2020,wang_prb_2021}. Hence, if there exists a 3DFOTI-to-3D-metal transition, we expect that $\langle\zeta_{W,L}(E)\rangle$ is independent of $L$ for $W<W_{c,2}$. This is indeed what we found as shown in Figs.~\ref{fig2}(c,d,e,f) where $W_{c,2}\simeq 3.0$ for cases (i) and (ii). Therefore, our numerical data confirm the existences of surface states for $W\in [W_{c,1},W_{c,2}]$.
\par

For stronger disorders $W>W_{c,2}$, the states of $E=0.02+i\gamma_0$ in Fig.~\ref{fig2}(c,d) are bulk states. Such states are extended states, and a third quantum phase transition from non-Hermitian diffusive metals to Anderson insulators, i.e., the Anderson localization transitions, happens at a higher critical disorder $W_{c,3}$. This can be seen from the finite-size scaling analysis of participation ratio $p_2$ for a given energy $E$, which satisfies a single-parameter scaling function near $W_{c,3}$~\cite{pixley_prl_2015,wang_prb_2020}
\begin{equation}
\begin{gathered}
p_2(W)=L^D [f(L/\xi)+CL^{-y}]
\end{gathered}\label{eq3_3}
\end{equation}
In Eq.~\eqref{eq3_3}, $D$ is the fractal dimension of the critical wave function, $f(x=L/\xi)$ is the unknown scaling function, $C$ is a constant, and $y>0$ is the exponent of the irrelevant variable. $\xi$ is the correlation length and diverges as $\xi\sim |W-W_{c,3}|^{-\nu}$ with $\nu$ being the critical exponent characterized the universality class of Anderson localization transitions. The identification of an Anderson localization transition is guided by the following observations: (i) For extended (localized) states, the quantity $Y_L(W)=p_2 L^D-CL^{-y}$ increases (decreases) with system size $L$. (ii) Near $W_{c,3}$, $Y_L$ collapses to a two-branch smooth function, i.e., $f(x)$.
\par

\begin{figure}[htbp]
\centering
\includegraphics[width=0.45\textwidth]{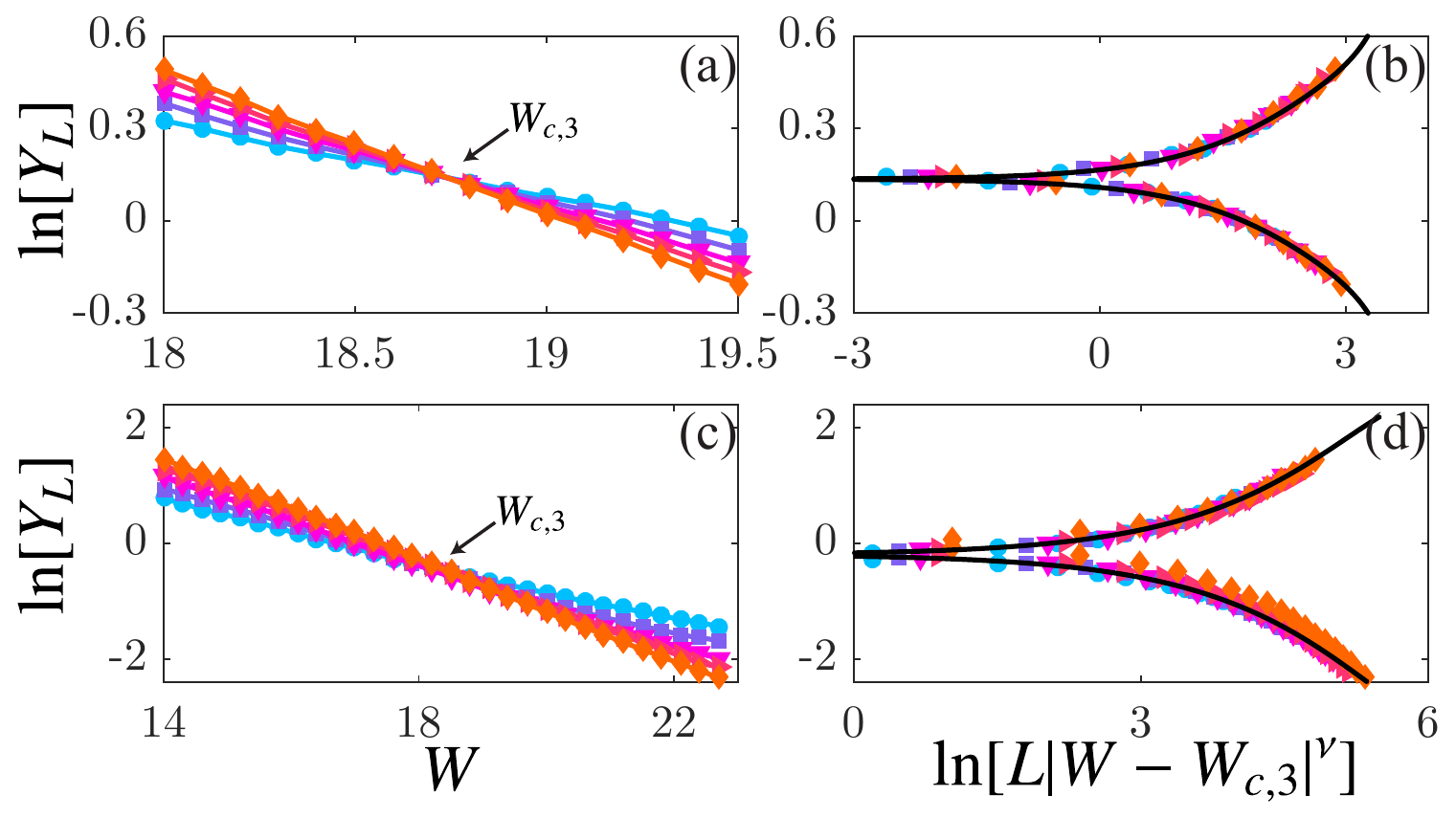}
\caption{(a) $\ln [Y_L]$ as a function of $W$ for $\gamma_0=-0.01,\gamma_{1,2,3}=0$ and $E=0.02+i\gamma_0$. The system sizes are $L=12$ (circles), 16 (squares), 20 (down triangles), 24 (right triangles), and 28 (diamonds). (b) Scaling function $\ln [Y_L]=\ln [f(x=\ln[L|W-W_{c,3}|^{\nu}])]$ for the Anderson localization transition in (a). (c,d) Same as (a,b) but for $\gamma_1=-0.01,\gamma_{0,2,3}=0$. The arrows point to $W_{c,3}$. Each point is average over more than $10^3$ ensembles.}
\label{fig3}
\end{figure}

Numerical evidence of Anderson localization transitions is displayed in Figs.~\ref{fig3}(a,c) for (i) $\gamma_0=-0.01,\gamma_{1,2,3}=0$ and (ii) $\gamma_{1}=0.01,\gamma_{0,2,3}=0$, respectively. Clearly, there exist the crossing points at the critical disorders $W_{c,3}=18.73\pm 0.02$ for case (i) and $W_{c,3}=17.4\pm 0.1$ for case (ii), where $Y_L(W)$ are independent of system sizes. Data of $d\ln [Y_L]/dL$ is positive (negative) for $W<W_{c,3}$ ($W>W_{c,3}$), indicating the non-Hermitian system is a diffusive metal (Anderson insulator). $\ln [Y_L(x=\ln [L|W-W_{c,3}|^{\nu}])]$ near $W_{c,3}$ merge to two branches of one smooth scaling function, see Figs.~\ref{fig3}(b,d). 
\par

The finite-size scaling analysis yields $\nu=1.47\pm 0.03$, $D=1.73 \pm0.06$, $C=1.2\pm 0.1$, $y=0.7\pm 0.1$ for case (i) and $\nu=1.2\pm 0.1$, $D=2.03 \pm0.05$, $C=0.8\pm 0.2$, $y=1.4\pm 0.2$ for case (ii). The critical exponent $\nu$ of case (i) is identical to that of Anderson localization transitions of its Hermitian counterpart~\cite{wang_prresearch_2020}, indicating that the appearance of the non-Hermitian potential $\sum_{\bm{i}} c^\dagger_{\bm{i}} (i\gamma_0\Gamma^0) c_{\bm{i}}$ does not change the universality class. On the other hand, the critical exponent of case (ii), say $\nu=1.2\pm 0.1$, is very close to that of symmetry class AI$^\dagger$ in the non-Hermitian AZ classification of the non-Hermitian systems~\cite{xluo_prresearch_2022}. 
\par

\begin{figure}[htbp]
\centering
\includegraphics[width=0.45\textwidth]{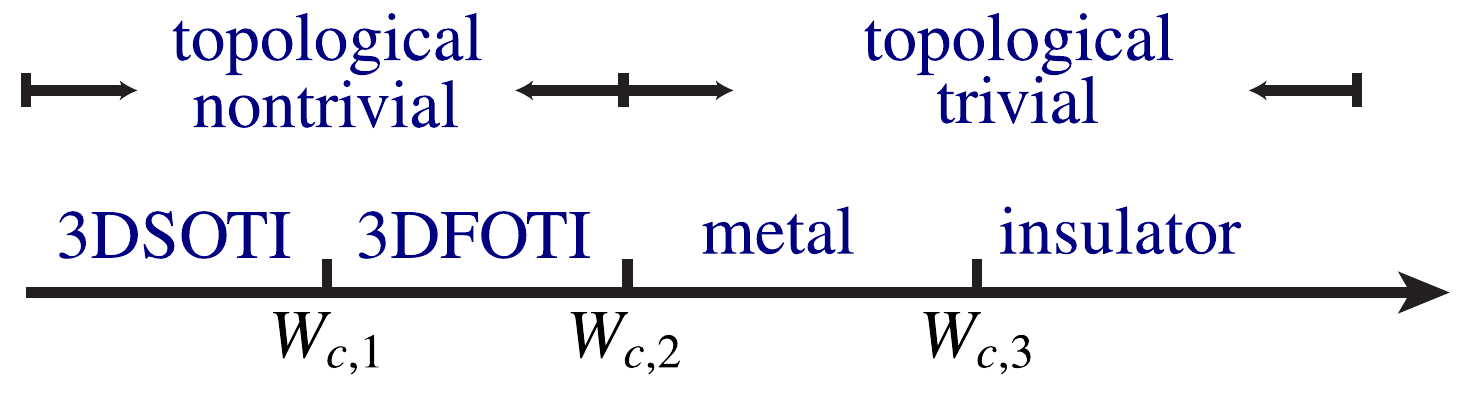}
\caption{A schematic route of disorder-induced quantum phase transitions of non-Hermitian 3DSOTIs. With increasing the disorder $W$, transitions from 3DSOTIs to 3DFOTIs then to topologically trivial metals happen at $W_{c,1}$ and $W_{c,2}$, respectively. Then, an Anderson localization transition will happen at a higher disorder $W_{c,3}$.}
\label{fig4}
\end{figure}

Through numerical analysis of $\langle\eta_{W,L}(E)\rangle$ and $\langle \xi_{W,L}(E)\rangle$, we have proven that a non-Hermitian 3DSOTI undergoes transitions to a 3DFOTI at $W_{c,1}$ then to a topological trivial system at $W_{c,2}$. We further substantiate that the topological trivial system is a diffusive metal in $[W_{c,2},W_{c,3}]$ and becomes an Anderson insulator for $W>W_{c,3}$ through the finite-size scaling analysis of participation ratios $p_2$. Our numerical results thus suggest a general route of quantum phase transitions of non-Hermitian 3DSOTIs, see Fig.~\ref{fig4}, which is very similar to that of Hermitian 3DSOTIs~\cite{wang_prresearch_2020}. 
\par

\section{Transmission coefficients}
\label{sec4}

\begin{figure}[htbp]
\centering
\includegraphics[width=0.45\textwidth]{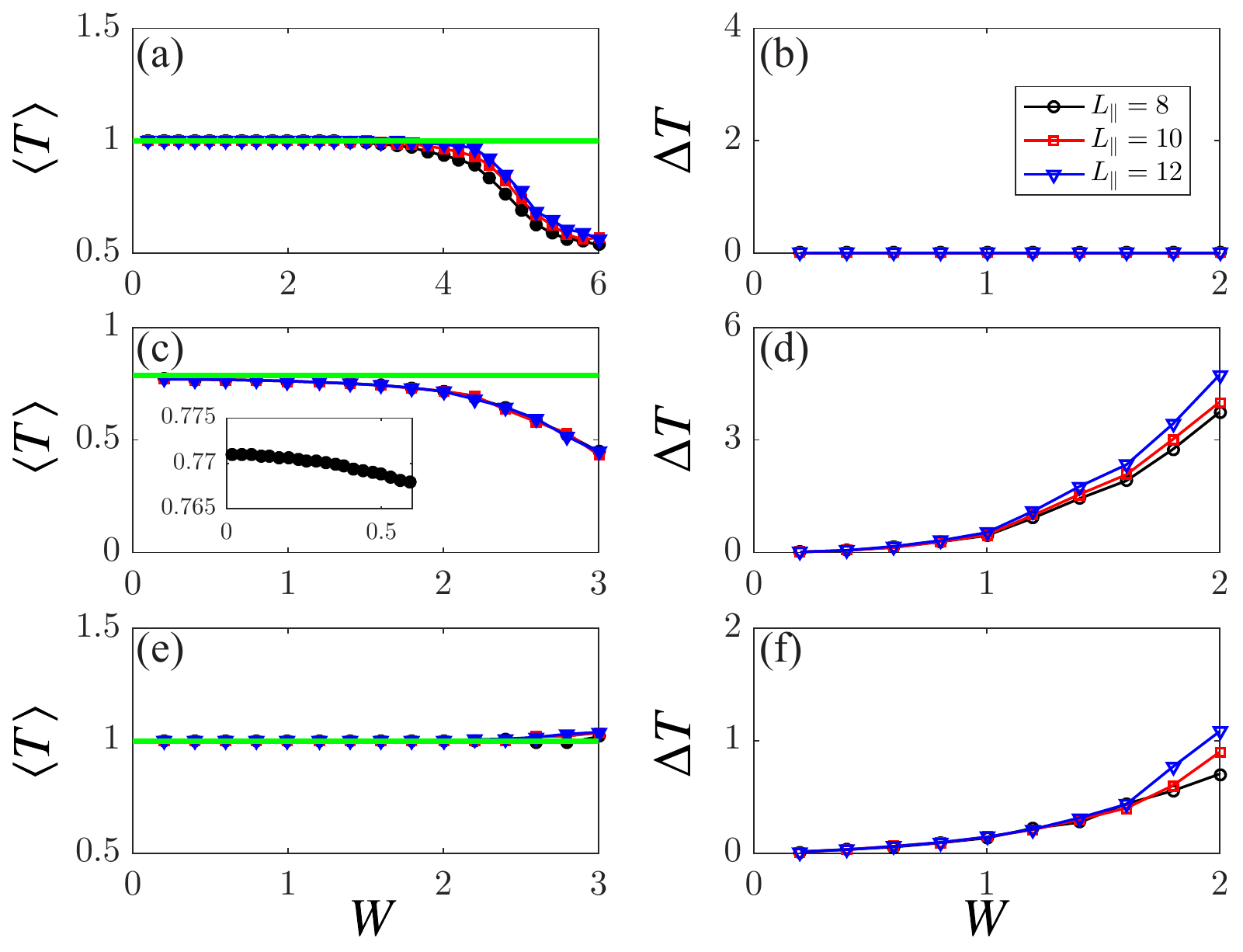}
\caption{(a,b) $\langle T \rangle$ and $\Delta T$ as a function of $W$ for various $L_\parallel$ for $\gamma_{0,1,2,3}=0$ (Hermitian 3DSOTIs). Here, $L_\perp=12$. (c,d) Same as (a,b) for $\gamma_{0}=-0.01,\gamma_{1,2,3}=0$. (e,f) Same as (a,b) for $\gamma_{1}=-0.01,\gamma_{0,1,2}=0$. Green lines in (a,b,c) are Eq.~\eqref{eq4_1}. Insert in (c) is an zoom-in picture in the small-disorder regime.}
\label{fig5}
\end{figure}

Having determined the phase diagram of disordered non-Hermitian 3DSOTIs, we are now in the position to investigate the quantum transport of hinge states. We analyse the transmission coefficient of a Hall bar of size $L_\parallel/\sqrt{2}\times L_\parallel/\sqrt{2}\times L_\perp$ connected to two semi-infinite leads at two ends along the $z-$direction. Particles in the Hall bar are govern by the Hamiltonian of Eq.~\eqref{eq2_1}, and the leads are clean and Hermitian, i.e., Eq.~\eqref{eq2_1} without the disorders and the $\gamma$-terms. Some previous works have studied the conductivities of non-Hermitian Chern insulators within the framework of Kubo-Greenwood formalism~\cite{philip_prb_2018,chen_prb_2018,groenendijk_prr_2021} by assuming conductor together with leads remains quantum mechanically coherent. This approach neglects the intrinsic loss of quantum coherence due to the non-Hermicities in the self-energy. Here, we use B\"{u}ttiker's approach combined with the non-equilibrium Green function method (NEGF) to calculate the transmission coefficients~\cite{datta,buttiker_prb_1986,meir_prl_1992,jiang_prl_2009}, where a virtual lead of zero net particle flux is introduced to model the incoherent processes that are not in the usual Kubo-Greenwood formalism.
\par

Figure~\ref{fig5} shows some typical results of mean transmission coefficients $\langle T\rangle$ 
for $E=0.02+i\gamma_0$. In general, transmission coefficients for the $\pm z$-moving fluxes are 
different in disordered non-Hermitian systems~\cite{luo_prl_2021}. While, for the hinge channels 
governed by Eq.~\eqref{eq2_3}, the transpose symmetry is protected, and disorder-average 
transmission coefficients for the $\pm z$-moving fluxes are the same. Without the non-Hermitian 
potentials, the mean transmission coefficient is quantized at 1, i.e., $\langle T \rangle=1$, and the 
fluctuations of transmission coefficients, defined as $\Delta T=(\langle T^2\rangle-\langle T\rangle^2)^{1/2}$, 
are vanishingly small ($\Delta T\sim 10^{-7}$ in our numerical calculations) until $W_{c,1}= 2.6$, see 
Fig.~\ref{fig5}(a,b). This is a standard feature of chiral hinge states in disordered Hermitian 
3DSOTIs~\cite{wang_prresearch_2020}. 
\par

In the presence of a non-local loss $\gamma_0=-0.01,\gamma_{1,2,3}=0$, $\langle T\rangle$ is not quantized any more even for $W<W_{c,1}$ where the hinge states persist. This can be seen by Fig.~\ref{fig5}(c) and its inset. As shown in Fig.~\ref{fig5}(d), the fluctuations of transmission coefficients are also non-zero ($\Delta T\sim 10^{-3}$) and at least four order of magnitude larger than those of Hermitian cases. In addition, $\Delta T$ increases with $W$ and the system sizes $L$, indicating that the fluctuation of transmission coefficients is not due to numerical errors and cannot be eliminated in the thermodynamic limit where $L\to\infty$. 
\par

Differently, for Hermitian hinge states in non-Hermitian 3DSOTIs where $\gamma_1=-0.01$ and $\gamma_{0,2,3}=0$, the mean transmission coefficient $\langle T\rangle$ is still equal to the number of chiral hinge channel (here is 1). The values of mean transmission coefficients in Fig.~\ref{fig2}(e) depend neither on the width of the Hall bar nor on disorders in $W<W_{c,1}$, a behavior which is strongly reminiscent of the Hermitian chiral hinge channels [see Fig.~\ref{fig2}(a)]. The plateau of $\langle T\rangle=1$ can also be observed for $\gamma_{2}=-0.01,\gamma_{0,1,3}=0$ (not shown here). The akin results have been observed in non-Hermitian 3DSOTIs and other non-Hermitian topological systems like Chern insulators and topological Anderson insulators in our early work~\cite{cwang_prb_2022}. However, the fluctuations of transmission coefficients are always non-zero. Similar to the non-Hermitian chiral hinge states, the fluctuations of transmission coefficients are not due to numerical error and increase with the system sizes, see Fig.~\ref{fig5}(f).
\par

Similar behaviors of the mean transmission coefficient $\langle T\rangle$ and the fluctuation of transmission coefficient $\Delta T$ have been observed for non-Hermitian 3DSOTIs with different disorders. In Appendix~\ref{sec_disorder}, we have considered two additional types of disorders: ``non-Hermitian'' on-site disorder and non-uniform in the subspace of interior degree of freedom. Again, we find that $\langle T\rangle$ equals 1 if the chiral hinge states are Hermitian; otherwise, $\langle T\rangle\neq 1$. In both cases, $\Delta T$ is finite. 
\par

A fairly simple theory, where the incoherent processes are artificially encoded in the particle density, can explain the mean transmission coefficient $\langle T\rangle$ displayed in Figs.~\ref{fig5}(a,c,e). Within the framework of NEGF, the real parts of energy spectra $\text{Re}[E]$ are interpreted as the quasi-particle energies, and the imaginary parts $\text{Im}[E]$ associate with incoherent processes. Take electrons as examples. The second law of thermodynamics requires $\text{Im}[E]\leq 0$ such that one can treat $\hbar/(2|\text{Im}[E]|)$ as the average relaxation time of an electron. Without disorders, the current of a coherent hinge mode with momentum $k$ is given by $I_k=\rho_k v_k f(E_k)$ with the electron density $\rho_k$,  the Fermi-Dirac distribution $f(E_k)$, and the group velocity $v_k=\partial_k E_k/\hbar$~\cite{hu_prl_2013}. In the presence of non-Hermitian potentials, we expect that the electron density is reduced by a factor of $\exp[-2|\text{Im}[E]|L_\perp/t]$ due to the relaxation of electrons.  
\par

Consider two reflectionless leads with chemical potentials $\mu_1$ and $\mu_2$, 
respectively, and $|\mu_{1,2}|<2\Delta$ with $2\Delta$ being the gap of the energies 
of hinge states. At the zero temperature, we obtain the net current flow of 
$I=(e/h)(\mu_1-\mu_2)e^{-2|\text{Im}[E]|L_\perp/t}$ and the dimensionless conductance, 
which is equal to the transmission coefficient according to the Landauer formula~\cite{datta}, 
\begin{equation}
\begin{gathered}
g=T=\exp[-2|\text{Im}[E]|L_\perp/t],
\end{gathered}\label{eq4_1}
\end{equation}
see Appendix~\ref{sec_clean} for more details. Therefore, for a non-local loss 
$\gamma_0<0$ and $\gamma_{1,2,3}=0$ where $|\text{Im}[E]|=|\gamma_0|>0$, the 
coherent propagation of an electron in the hinge channel is destroyed when it interacts 
with the surroundings, which leads to $T<1$. Differently, for $\gamma_0=\gamma_3=0$ 
and $\gamma_{1,2}\neq 0$, the hinge channels are Hermitian with real energy spectra 
[see Eq.~\eqref{eq2_3}] such that it retains its identity even at lead and $T=1$. Even 
though Eq.~\eqref{eq4_1} is derived for $W=0$, we find it still works well for 
$\langle T\rangle$ at weak disorders [see the solid lines in Figs.~\ref{fig2}(a,c,e) without 
any fitting parameter].
\par

\begin{figure}[htbp]
\centering
\includegraphics[width=0.45\textwidth]{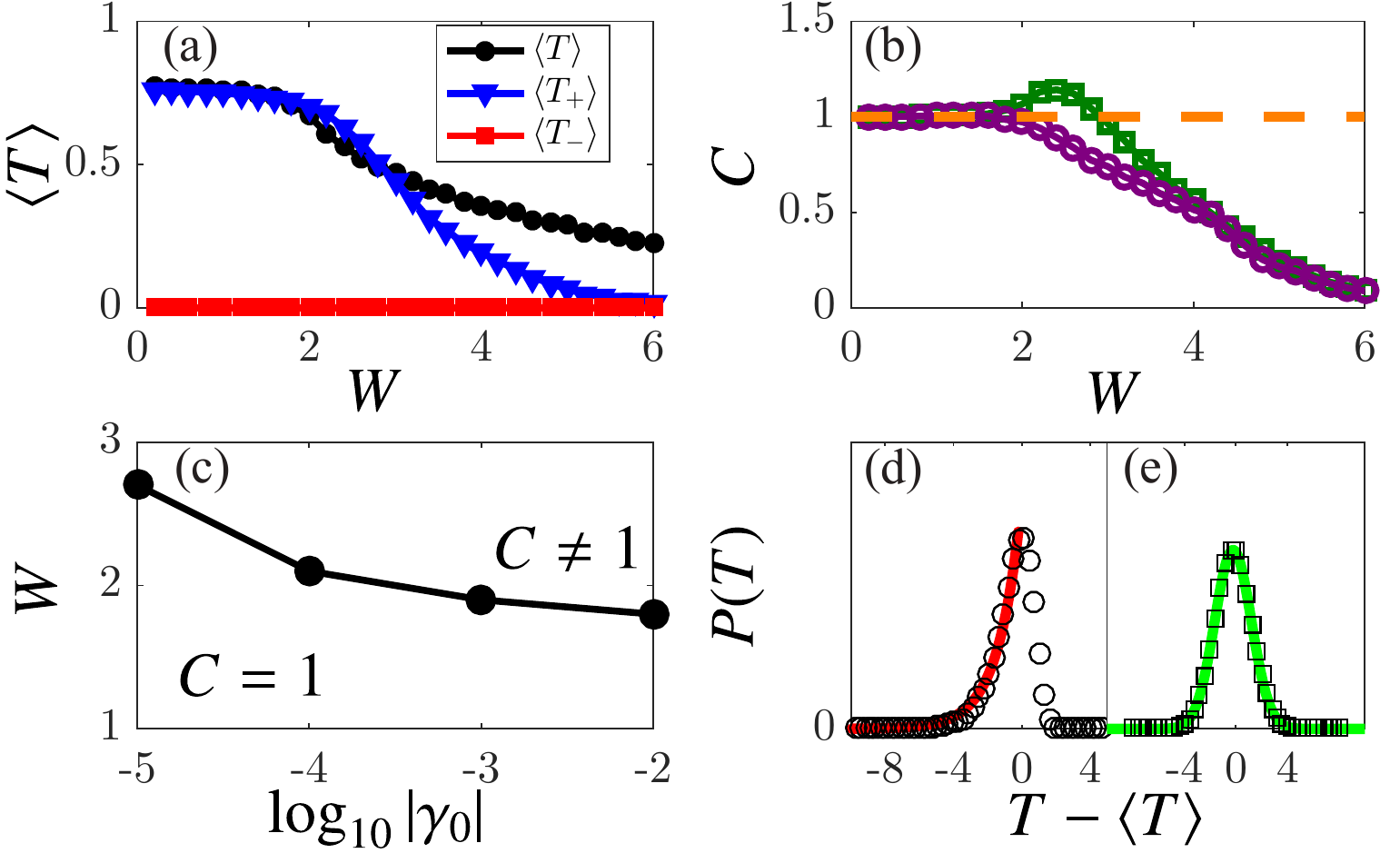}
\caption{(a) $\langle T\rangle,\langle T_+\rangle,\langle T_-\rangle$ v.s $W$ for 
$\gamma_0=-0.01,\gamma_{1,2,3}=0$, $L_\parallel=8$ and $L_\perp=12$. 
(b) $C$ as a function of $W$ for $\gamma_0=-0.01,\gamma_{1,2,3}=0$ (green) 
and $\gamma_1=-0.01,\gamma_{0,2,3}=0$ (purple). Orange dash line locates $C=1$. 
(c) $W_{c,1}$ v.s $\log_{10}|\gamma_0|$ ($\gamma_0<0$) with $\gamma_{1,2,3}=0$. 
(d.e) Distribution of $T$ for (d) $\gamma_0=-0.01,\gamma_{1,2,3}=0$ and (e) 
$\gamma_1=-0.01,\gamma_{0,2,3}=0$ . The solid lines are the best fits to a one-sided 
exponential distribution $P(T)\sim \exp[-\lambda |T-\langle T\rangle|]$ (red) and a Gaussian 
distribution (green). Here, $W=0.2$, and $L_\parallel=L_\perp=12$.}
\label{fig6}
\end{figure}

Still, there is one crucial question: Why transmission coefficients of hinge channels 
fluctuate in non-Hermitian 3DSOTIs? In Hermitian systems, the fluctuations come 
from the random opening and closing of a conducting channel as disorder configurations 
vary due to the inevitable diffusion of defects and impurities at finite temperatures. 
Each conducting channel contributes exactly one quantum conductance no matter what 
its dispersion is. Thus, there is no fluctuation of the transmission coefficients in 
one-dimensional chiral hinge channels as long as disorders do not destroy the channel. 
One may attribute the fluctuation in transmission coefficients to the broken chirality of 
hinge states due to the non-Hermitian potentials. 
\par

To test the correctness of this conjecture, we add a local loss $-i\delta \Gamma^0$ for $y>0$ ($y<0$) of a very large $\delta=t$, i.e., particles in one hinge $x=0,y=L_\parallel/2$ ($x=0,y=-L_\parallel/2$) completely decay, and calculate the corresponding mean transmission coefficient $\langle T_{+}\rangle$ ($\langle T_{-}\rangle$). The chirality of the hinge state can be defined as
\begin{equation}
\begin{gathered}
C=(\langle T_+\rangle-\langle T_-\rangle)/\langle T\rangle.
\end{gathered}\label{eq4_2}
\end{equation}
If the chirality is preserved, say a up-moving current flux state propagating in one hinge $x=0,y=L_\parallel/2$ without backward scattering, we expect $\langle T_+\rangle=\langle T\rangle$, $\langle T_-\rangle=0$, and $C=1$. Otherwise, $C\neq 1$ if the chirality is broken. 
\par

Figure~\ref{fig6}(a) shows $\langle T\rangle$, $\langle T_+\rangle$, and $\langle T_-\rangle$ as a function of disorder strengths $W$ for $\gamma_0=-0.01,\gamma_{1,2,3}=0$. Clearly, $\langle T_-\rangle$ vanishes (about $10^{-11}$) due to the strong loss, while $\langle T \rangle=\langle T_+\rangle$ for $W<W_{c,1}\simeq 1.8$ [the same critical disorder obtained by Fig.~\ref{fig2}(a)], indicating no backward scattering of particles in such hinge channel. This can be also seen from Fig.~\ref{fig6}(b): $C$ is always quantized as long as $W<W_{c,1}$, no matter what kind of non-Hermicities is applied. Therefore, the quantized $C$ should be convincing evidence of the preservation of chirality. 
\par

The quantized chirality $C=\pm 1$ (the sign depends on $B$) behaves as a 
fingerprint of chiral hinge states in non-Hermitian disordered 3DSOTIs, similar 
to the quantized quadrupole for the corner states in Hermitian topological 
insulators~\cite{benalcazar_science_2017}. Specifically, we utilize the quantized 
$C=1$ to identify the non-Hermitian 3DSOTI phase in the $\log_{10}[|\gamma_0|]-W$ 
plane numerically, see Fig.~\ref{fig6}(c). Our numerical data suggest the critical 
disorder $W_{c,1}$ decreases with $|\gamma_0|$.     
\par

Although the chirality of hinge states is preserved and the backward scattering is absent, the fluctuation of transmission coefficients can still arise from the scattering of the non-Hermitian potentials. This can be seen from a general formulation of a two-terminal multi-channels conductor in the beautiful book by Datta \cite{datta}:
\begin{equation}
\begin{gathered}
R_{11}+T_{12}+\text{Tr}[\tilde{\Gamma}_1 G^R[i(H-H^\dagger)]G^A]=M_1
\end{gathered}\label{eq4_3}
\end{equation}
with $M_1$ being the number of conducting channels of lead 1. $R_{11}$ is the reflection coefficient, and $T_{12}$ is the transmission coefficient from lead 2 to 1. $\tilde{\Gamma}_{1(2)}=i(\Sigma^{R}_{1(2)}-\Sigma^{A}_{1(2)})$, $G^{R}=[E-H-\Sigma^R]^{-1}$ and $G^A=[E-H^\dagger-\Sigma^A]^{-1}$ are the retarded and advanced Green's functions with $H$ being the Hamiltonian of the conductor and $\Sigma^{R(A)}$ being the total retarded (advanced) self-energy, respectively. $\text{Tr}[\tilde{\Gamma}_1 G^R[i(H-H^\dagger)]G^A]$ is the transmission coefficient for a particle incident in the virtual lead to reach the lead 1. Therefore, one cannot self-consistently compute the transmission coefficient $T_{12}$ as if there were no incoherent processes (virtual lead) since it is affected by these events. Following B\"{u}ttiker's approach, we obtain the transmission coefficient of the chiral hinge states
\begin{equation}
\begin{gathered}
T=T_{12}=(1-R_{11})-\text{Tr}[\tilde{\Gamma}_1 G^R[i(H-H^\dagger)]G^A],
\end{gathered}\label{eq4_4}
\end{equation}
see Appendix~\ref{sec_butt} for more details. Here, $M_1=1$ since we consider the chiral hinge transport. The contributions from the chiral hinge states and the scattering by non-Hermitian potentials are separated in the first and the second terms of Eq.~\eqref{eq4_4}. Therefore, even though no fluctuation is in the first term due to the preservation of chirality (no backward scattering means $R_{11}=0$), scattering by non-Hermicities always leads to fluctuations of transmission coefficients since it relates to the bulk Hamiltonian $H$ rather than the topological boundary states. Such fluctuations increase with $i(H-H^\dagger)$ that is proportional to the strength of non-Hermicities. 
\par

As the fluctuations of transmission coefficients of chiral hinge states in non-Hermitian 3DSOTIs are inherent, the understanding of their distributions $P(T)$ should be very interesting. Numerically, we find $P(T)$ of metallic (bulk and surface) and localized states of Eq.~\eqref{eq2_2} follow the Gaussian and the log-normal distributions, respectively, similar to its Hermitian counterparts~\cite{beenakker_rmp_1997}. While, for $P(T)$ of chiral hinge states, their Hermiticities play a significant role again. For Hermitian chiral hinge states, $P(T)$ is also Gaussian. However, it becomes very asymmetric with a long tail towards small $T$ ($T<\langle T\rangle$) if the chiral hinge states are dissipative, see Fig.~\ref{fig6}(d). Exactly, the distributions always evolve to a one-sided exponential distribution $P(T)\sim\exp[-\lambda|T-\langle T\rangle|]$, where the coefficient $\lambda$ depends on the disorders and the non-Hermitian potentials.
\par

\section{Discussions}
\label{sec5}

We would like to discuss the experimental relevance of our theory before the conclusion. 
Loss can be introduced by the relaxation time of electrons such that results of $\gamma_0< 0$ 
and $\gamma_{1,2,3}=0$ can be directly applied to the electronic 3DSOTIs, e.g., 
Bi$_{2-x}$Sm$_x$Se$_3$~\cite{yue_natphys_2019}. However, gains are not allowed for 
electronic systems in principle. We thus suggest the chiral hinge magnons as a potential 
platform for observing the reported Hermitian chiral hinge transport, e.g., 
$\gamma_1\neq 0$ and $\gamma_{0,2,3}=0$. The realization of chiral hinge magnons has 
been illustrated in Ref.~\cite{amook_prb_2021}. The loss of magnons, known as the Gilbert 
damping, is the intrinsic property of spin systems~\cite{yang_prl_2018}, while, recently, it has 
been shown that the gain (the negative Gilbert damping) can be realized by the magneto-optical 
interaction and artificially tuned~\cite{cao_prb_2022}. In real magnetic systems, disorders 
exist ubiquitously. In Ref.~\cite{cwang_prb_2018}, it is shown that the forms of randomness 
used in this work are good approximations of disorders and can be used to explain the origin 
of the well-known spin Seebeck effect. Instead of magnonic chiral hinge states, all reported 
results are nevertheless appropriate for phononic~\cite{susstrunk_science_2015,li_natphys_2018,garcia_nature_2018} 
and photonic~\cite{khanikaev_natmater_2013} topological insulators, where gains are allowed.
\par

Besides, Hermitian chiral boundary (edge or hinge) states can appear in a large family of 
topological phases, including Chern insulators, 3DSOTIs, and topological Anderson 
insulators~\cite{cwang_prb_2022}. The physics discussed here is applicable to other 
non-Hermitian topological phases with Hermitian chiral edge states. Stimulated by this, 
we propose another system, the 2D laser cavity network on a honeycomb lattice that 
supports chiral edge states, as an easy-to-access platform to test our predictions. 
We show the flexibility of this proposal in Appendix~\ref{sec_laser}.
\par

\section{Conclusion}
\label{sec6}

In conclusion, chiral hinge transport in disordered non-Hermitian 3DSOTIs is studied numerically. Chiral hinge states can be either Hermitian or non-Hermitian, depending on the form of the non-Hermitian potentials. Both can survive at finite disorders and undergo the transitions to surface states when increasing the disorder up to a critical value $W_{c,1}$. The mean transmission coefficients are equal and nonequal to the number of chiral channels for Hermitian and non-Hermitian chiral hinge states. The transmission coefficients always fluctuate whether the chiral hinge states are Hermitian or non-Hermitian. The fluctuations of transmission coefficients come from the scatterings of non-Hermitian potentials that describe incoherent processes involving the interactions between particles and their surroundings. Furthermore, the Hermiticity of chiral hinge states also affects the distribution of transmission coefficients. The distribution of Hermitian chiral hinge states is Gaussian; otherwise, it is a one-sided exponential distribution.  
\par

\begin{acknowledgments}

This work is supported by the National Key Research and Development Program of China 2020YFA0309600, the National Natural Science Foundation of China (Grants No.~11704061 and No.~11974296), and Hong Kong RGC (Grants Nos.~16301518 16301619, and 16302321).

\end{acknowledgments}

\appendix

\section{Interpretation of 3DSOTI-to-3DFOTI transitions}
\label{sec_transitions}

\begin{figure*}[htbp]
\centering
\includegraphics[width=0.75\textwidth]{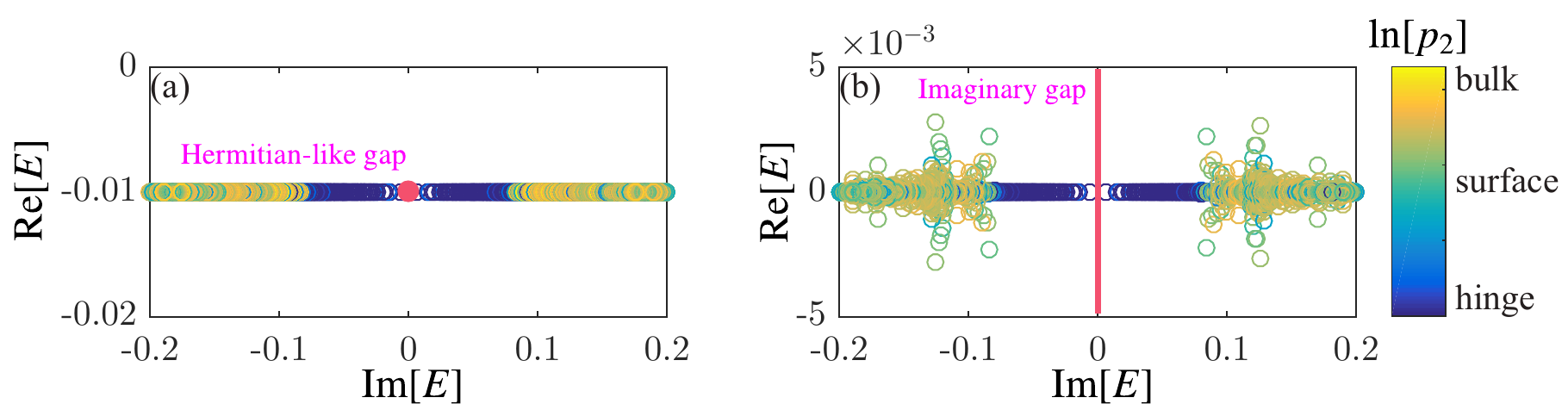}
\caption{Complex-energy spectra of $H$ with (a) $\gamma_0=-0.01,\gamma_{1,2,3}=0$ and (b) $\gamma_1=-0.01,\gamma_{0,2,3}=0$. Here, $W=0.01$, $L_\parallel=L_\perp=30$. The other parameters are the same as those Fig.~\ref{fig1}. Only a part of complex energies (near the regime of chiral hinge states) are displayed. Colors map $\ln[p_2]$ with $p_2$ being the participation ratios, from which one can easily identify the chiral hinge states (blue). Colorbar is the same as that in Fig.~\ref{fig1}.}
\label{figa1}
\end{figure*}

In Sec.~\ref{sec3}, we argue that the 3DSOTI-to-3DFOTI transitions happen when the gap of topological surface states closes. To prove our reasoning and analysis, a non-Hermitian generalization of the concept of the energy gap is necessary. 
\begin{itemize}
\item \emph{Case (i): $\gamma_0=-0.01,\gamma_{1,2,3}=0$}.\par
In this situation, the generalization of energy gap is trivial since the complex-energy spectra can be mapped to a real-energy spectra by shifting a constant $-i\gamma_0$. Such gap is ``Hermitian-like'', as shown in Fig.~\ref{figa1}(a) that displays $\text{Re}[E]$ with respect to $\text{Im}[E]$ near the regime where chiral hinge and surface states appear. For their Hermitian counterpart, the mechanism of the topological phase transitions has been elucidated in our early work~\cite{wang_prresearch_2020}: The phase transition from a 3DSOTI to a 3DFOTI happens when the real-energy gap closes. 
\item  \emph{Case (ii): $\gamma_1=-0.01,\gamma_{0,2,3}=0$}.\par
In this case, the definition of an energy gap is highly non-trivial. We follow the definition given in  Ref.~\cite{kawabata_prx_2019}: The complex-energy band of $H$ of $\gamma_1=-0.01,\gamma_{0,2,3}=0$ does not cross a reference line in the complex-energy plane (here it is the $y$-axis) as shown in Fig.~\ref{figa1}(b). The quantum phase transition from 3DSOTIs to 3DFOTI if and only if the energy gap on the hinge closes at a critical disorder. 
\end{itemize}
\par

\begin{figure}[htbp]
\centering
\includegraphics[width=0.45\textwidth]{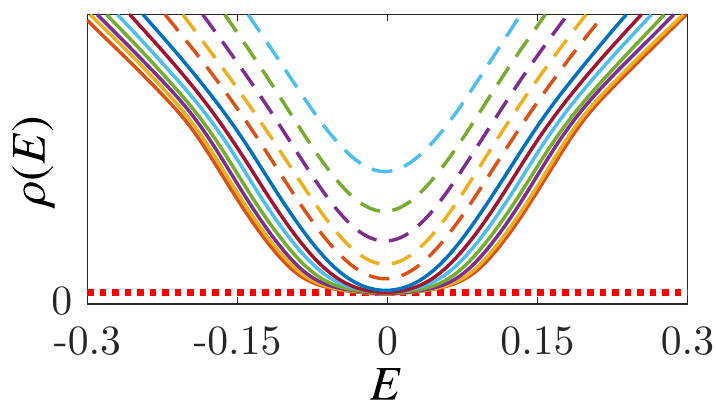}
\caption{$\rho(E)$ (arbitrary unit) of $\tilde{H}=H-i\gamma_0\Gamma^0$ for $\gamma_{0}=-0.01,\gamma_{1,2,3}=0$ for the windows of $[-0.3,0.3]$ and $L=100$. Disorder strengths are $W=0.7,0.9,\cdots,1.7$ (solid, from down to up) and $W=1.9,2.1,\cdots,2.7$ (dashed, from down to up). The dotted line shows the nonzero constant $\rho(E)$. Those data below (above) the critical disorder $W_{c,1}$ are plotted by the solid (dashed) lines. $10^2$ samples are used to average $\rho(E)$.}
\label{figa2}
\end{figure}

We can partially verify the above argument for the cases of Hermitian-like gaps by calculating the density of states (DOS) of the Hermitian part of $H$, say $\tilde{H}=H-i\gamma_0\Gamma^0$. In the absence of disorder $W=0$, the dispersion of the chiral hinge states in $\tilde{H}$ is linear in the crystal momentum $p$, see Eq.~\eqref{eq2_3}. Its contribution to the DOS is a constant, i.e., $\rho_{\text{hinge}}=1/t$. The feature of the constant DOS persists even in the presence of disorders, see Fig.~\ref{figa2}, which shows the average DOS, defined by $\rho(E)=\langle \sum_i \delta(E-E_i) \rangle/V$, for various disorders obtained by the well-established kernel polynomial method with $V$ being the volume of the disordered 3DSOTI. Apparently, the width of constant DOS becomes smaller with the increases of $W$, i.e., disorders tend to reduce the gap of surface states. At a critical disorder $W_{c,1}\simeq 1.8$, the gap of the surface states close (the plateau of constant DOS disappears), and a transition from 3DSOTIs to 3DFOTIs happens. The critical disorder is consistent with that obtained in Fig.~\ref{fig2}(c). 
\par

However, we currently lack a proper numerical method to calculate the ``DOS'' of the non-Hermitian 3DSOTI of the line gap shown in Fig.~\ref{figa1}(b). Hence, to illustrate how disorders close such line gap is beyond our ability, and, as far as we know, such a problem is unsolved yet and should be an interesting issue that deserves a further study.
\par

\section{Different forms of disorders}
\label{sec_disorder}

The effect of disorders on non-Hermitian 3DSOTIs can be understood within the framework of self-consistent Born approximation, which claims that the effective Hamiltonian of a disordered system can be written as $h_{\text{eff}}(\bm{k})=h(\bm{k})+\Sigma$ with $h(\bm{k})$ being the Hamiltonian of clean systems (here, it is Eq.~\eqref{eq2_2}) and $\Sigma$ being the self-energy induced by disorders. In the continuum limit, the first-order self-energy is $\Sigma_1=(1/V)\int U(x) d\bm{x}=\langle U(x)\rangle$ with $U(x)$ being the disorder potential.
\par

In the main text, we consider one type of random on-site potential $V=\sum_{\bm{i}}c^\dagger_{\bm{i}}v_{\bm{i}}\Gamma^0 c_{\bm{i}}$ whose spatial average is zero such that $\Sigma_1=0$. In this section, we study the chiral hinge transport under two different forms of disorders. The first one reads 
\begin{equation}
\begin{gathered}
\tilde{V}_1=\sum_{\bm{i}}c^\dagger_{\bm{i}}(w_{\bm{i}}+iu_{\bm{i}})\Gamma^0 c_{\bm{i}}
\end{gathered}\label{disorder1}
\end{equation}
with $w_{\bm{i}}$ and $u_{\bm{i}}$ being non-negative real numbers distributing randomly and uniformly in the ranges of $[-W/2,W/2]$ and $[-W_0-W/2,-W_0+W/2]$ with $\langle w_{\bm{i}}\rangle=0$, $\langle u_{\bm{i}}\rangle=-W_0$, and $\langle w_{\bm{i}}w_{\bm{j}}\rangle=\langle u_{\bm{i}}u_{\bm{j}}\rangle=(W^2/12)\delta_{\bm{ij}}$. It is easy to find that $\Sigma_1=-iW_0\Gamma^0$. We choose $\gamma_1=-0.01$ and $\gamma_{0,2,3}=0$ such that the chiral hinge states are Hermitian in the absence of disorders. Thus, with disorders, the Hermiticity of chiral hinge states is preserved (broken) for $W_0=0$ ($W_0\neq 0$).
\par

\begin{figure}[htbp]
\centering
\includegraphics[width=0.48\textwidth]{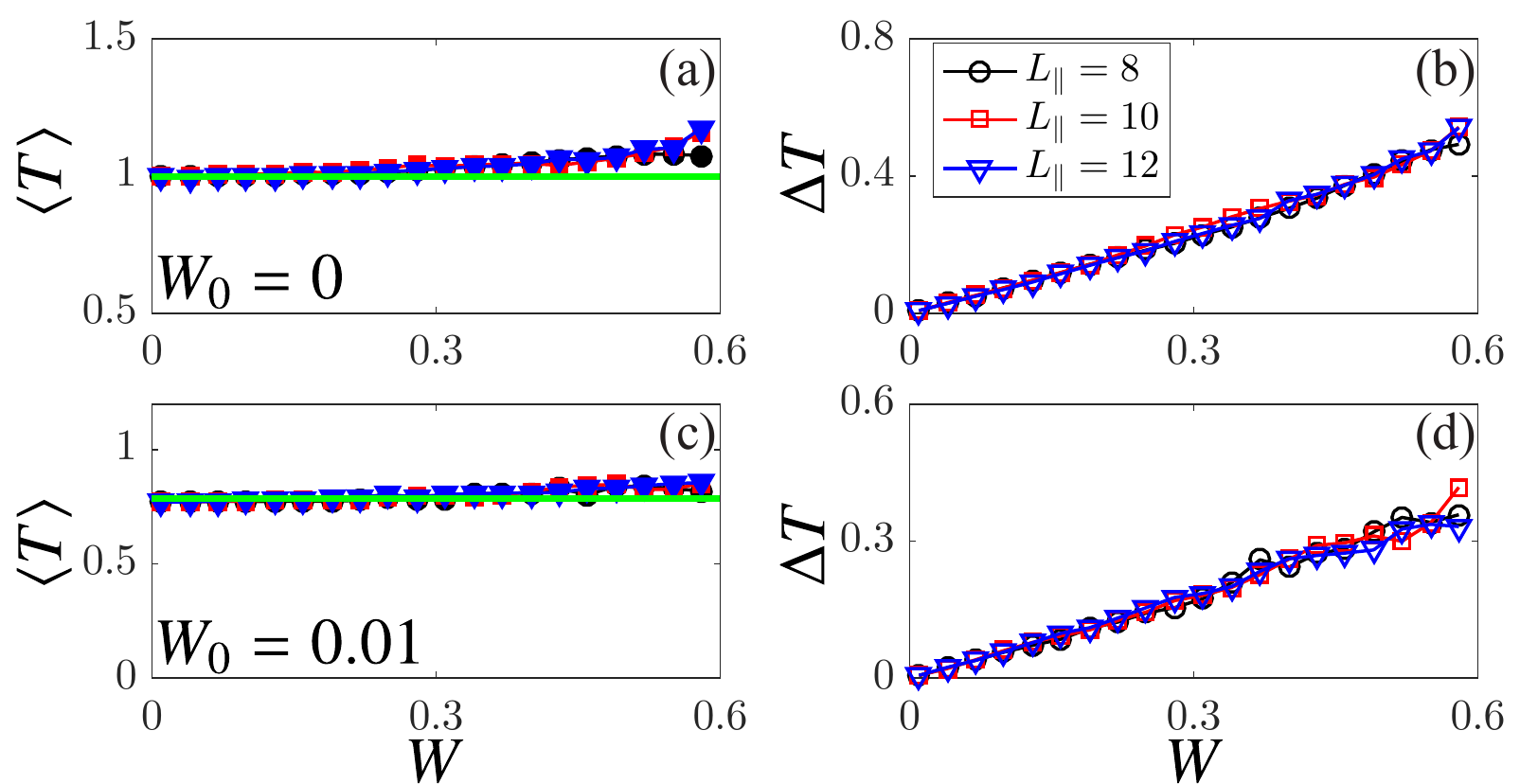}
\caption{(a,b) $\langle T\rangle$ (the filled symbols) and $\Delta T$ (the open symbols) as a function of $W$ for the disorders of Eq.~\eqref{disorder1} with $\gamma_1=-0.01$, $\gamma_{0,2,3}=0$, and $W_0=0$. Here, $L_{\perp}=12$, and $L_{\parallel}=8$ (the black squares), 10 (the red circles), 12 (the blue triangles). (c,d) Same as (a,b) but for $W_0=0.01$. The green solid lines are plotted by Eq.~\eqref{eq4_1}.}
\label{figb1}
\end{figure}

Results of $W_0=0$ and $W_0\neq 0$ are plotted in Fig.~\ref{figb1}. Clearly, for small disorders, $\langle T\rangle=1$ for $W_0=0$ but $\langle T\rangle<1$ for $W_0=0.01$. Therefore, no matter the disordered on-site potential is Hermitian or not, $\langle T\rangle=1$ ($\langle T\rangle\neq 1$) if the chiral hinge states are Hermitian (non-Hermitian). In both cases, $\langle T\rangle$ can be adequately described by Eq.~\eqref{eq4_1}. Besides, $\Delta T$ is finite and increases with $W$ for $W=0$ and 0.01. 
\par

It is worthwhile to mention that the average transition coefficient is larger than 1 in Fig.~\ref{figb1}(a). This is mainly due to the participation of bulk or surface states in transportation because strong disorder drives bulk and surface states close to the Fermi level such that one Hermitian hinge channel contributes one transmission coefficient, and partial contributions come from backward scattered bulk states.
\par
\quad\par

\begin{figure}[htbp]
\centering
\includegraphics[width=0.48\textwidth]{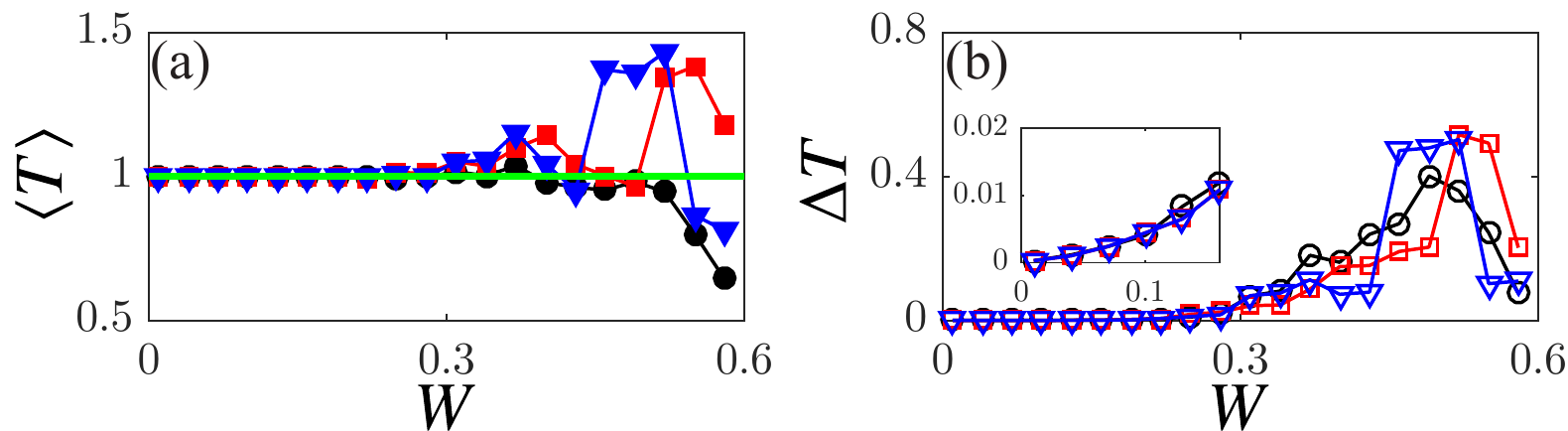}
\caption{(a,b) $\langle T\rangle$ (the filled symbols) and $\Delta T$ (the open symbols) v.s. $W$ for the disorders of Eq.~\eqref{disorder2} with $\gamma_1=-0.01$ and $\gamma_{0,2,3}=0$. Here, $L_{\perp}=12$, and $L_{\parallel}=8$ (the black squares), 10 (the red circles), 12 (the blue triangles). Insert: Zoom in picture in the small-disorders regime.}
\label{figb2}
\end{figure}

The second type of disorder reads
\begin{equation}
\begin{gathered}
\tilde{V}_2=\sum_{\bm{i}}c^\dagger_{\bm{i}}(w_{\bm{i}}\Gamma^0+u_{\bm{i}}\Gamma^1)c_{\bm{i}}
\end{gathered}\label{disorder2}
\end{equation}
with $\Gamma^1=s_0\otimes\sigma_3$ and the white noises $w_{\bm{i}}$ and $u_{\bm{i}}$ distributing uniformly in the range of $[-W/2,W/2]$. We interpret $s_{0,1,2,3}$ and $\sigma_{0,1,2,3}$ are the two-by-two unit matrices and Pauli matrices acting on the pseudo-spin and spin spaces, respectively. Hence, $\tilde{V}_2$ breaks the uniformity in the spin subspace but does not break the Hermiticity of the chiral hinge states. We then calculate $\langle T\rangle$ and $\Delta T$ as a function of $W$, see Fig.~\ref{figb2}. Evidently, for small disorders, $\langle T\rangle=1$ and $\Delta T\neq 0$.
\par

\section{A formula for the transmission coefficients}
\label{sec_clean}

In the clean limit, we have derived an expression of transmission coefficient $T$ in a fairly simple manner by using the low energy effective Hamiltonian for chiral hinge states [Eq.~\eqref{eq2_3}]. Let us take electrons as examples. The chiral hinge mode of a given $p_3=p$ carries a current: 
\begin{equation}
\begin{gathered}
I_p=\rho_p v_g f(\text{Re}[E]).
\end{gathered}\label{clean1}
\end{equation}
Here, $v_g$ is the group velocity. $\rho_p$ is the electron charge density that is related to the length of the hinge (here is $L_\perp$) and the travelling time $\tau_p$ of an electron moving from one lead to the other lead, i.e.,
\begin{equation}
\begin{gathered}
\rho_p=\dfrac{e}{L_\perp}\exp\left[ \dfrac{2\text{Im}[E]\tau_p}{\hbar} \right].
\end{gathered}\label{clean2}
\end{equation}
The effect of incoherent scattering is thus encoded in Eq.~\eqref{clean2}. For electrons, we require $\text{Im}[E]<0$ as explained in Sec.~\ref{sec4}. Without the non-Hermitian potentials, electrons propagate coherently, i.e., an electron injected from one lead will disappear in the other lead such that the electron density of a given mode is always $e/L_\perp$. Once the chiral hinge states are dissipative, a loss of electrons leads to $\rho_p< e/L_\perp$. The relaxation time reads $\tau_p=L_\perp/v_g$. For a general non-Hermitian Hamiltonian with a non-degenerate spectrum, we can define the standard left and right eigenkets as $H|u\rangle=E|u\rangle$ and $H|v\rangle=E^\ast|v\rangle$ with $\langle u|u\rangle=\langle v|v\rangle=1$. The group velocity can be defined as
\begin{equation}
\begin{gathered}
v_g=\dfrac{1}{\hbar}\dfrac{d}{dp}\text{Re}[\langle u|H|u\rangle]=\dfrac{1}{\hbar}\dfrac{d\text{Re}[E]}{dp}.
\end{gathered}\label{clean3}
\end{equation}
One can certainly use the terms like $\langle u|H|v\rangle$ with the bi-orthogonal metric $\langle u|v\rangle=1$ to define the group velocity, which, however, is the same as Eq.~\eqref{clean3}. The reason is that $v_g$ is solely determined by $\text{Re}[E]$ and is irrelevant to the metric. By using the effective Hamiltonian of chiral hinge states, we obtain $v_g=t/\hbar$. Then,
\begin{equation}
\begin{gathered}
I_p=\dfrac{e}{L_\perp}\exp\left[ \dfrac{-2|\text{Im}[E]|L_\perp}{t} \right]\dfrac{d\text{Re}[E]}{dk}.
\end{gathered}\label{clean4}
\end{equation}\par

Our idea is to put the effect of the incoherent process in the electron density by hand and to effectively treat the electron propagation as a coherent process. Consider two leads of chemical potentials $\mu_{1,2}$, respectively. Here, $|\mu_{1,2}|<\Delta$ with $2\Delta$ being the gap of the surface states where chiral hinge states exist. The current flowing along $+z$ direction thus reads
\begin{equation}
\begin{gathered}
I^{+}=\exp\left[ \dfrac{-2|\text{Im}[E]|L_\perp}{t} \right]\dfrac{e}{h}(\mu_1+\Delta),
\end{gathered}\label{clean5}
\end{equation}
where the Fermi-Dirac distribution function at the zero temperature is used. Likewise, the current flowing in the $-z$ direction is thus
\begin{equation}
\begin{gathered}
I^{-}=\exp\left[ \dfrac{-2|\text{Im}[E]|L_\perp}{t} \right]\dfrac{e}{h}(\mu_2+\Delta).
\end{gathered}\label{clean6}
\end{equation}
The net current is thus $I=I^+- I^-=e^{-2|\text{Im}[E]|L_\perp/t}(e/h)(\mu_1-\mu_2)$. Within the framework of the Landauer formula, the dimensionless conductance, defined as $g=Ih/[e(\mu_1-\mu_2)]$, is the same as the transmission coefficient $T$. Then, we obtain Eq.~\eqref{eq4_1}.
\par

\section{B\"{u}ttiker's approach}
\label{sec_butt}

We simply review B\"{u}ttiker's approach on calculating the transmission coefficients of a non-Hermitian system. In principle, the incoherent scattering processes are very difficult to handle. To overcome this barrier, B\"{u}ttiker has developed an approach by treating the incoherent scatterings as a virtual lead with a determined chemical potential~\cite{buttiker_prb_1986}. Consider the two-terminal setup with two real leads 1 and 2. Each lead has one propagating channel (chiral hinge channels). To model the incoherent scatterings, it is assumed that there is an extra virtual lead 3 with two channels 3 and 4. Conservation of particles requires that
\begin{equation}
\begin{gathered}
\left\{
\begin{array}{c}
R_{11}+T_{12}+T_{13}+T_{14}=1 \\
T_{21}+R_{22}+T_{23}+T_{24}=1 \\
T_{31}+T_{32}+R_{33}+T_{34}=1 \\
T_{41}+T_{42}+T_{43}+R_{44}=1 \\
\end{array}
\right..
\end{gathered}\label{butt1}
\end{equation}
Here, $R_{ii}$ is the reflection coefficient of a given channel $i$, and $T_{ij}$ is the transmission coefficient from the channel $i$ to the channel $j$. 
\par

Let us take electrons as examples. In the absence of a magnetic field, the transmission coefficient are symmetric, i.e., $T_{ij}=T_{ji}$. Then, the net current to the channel 1 is thus given by
\begin{equation}
\begin{gathered}
I_1=\dfrac{e}{h}\left[ (1-R_{11})\mu_1-T_{12}\mu_2-(T_{13}+T_{14})\mu_3 \right].
\end{gathered}\label{butt2}
\end{equation} 
Likewise, we have
\begin{equation}
\begin{gathered}
I_2=\dfrac{e}{h}\left[ (1-R_{22})\mu_2-T_{21}\mu_1-(T_{23}+T_{24})\mu_3 \right],
\end{gathered}\label{butt3}
\end{equation} 
\begin{equation}
\begin{gathered}
I_3=\dfrac{e}{h}\left[ (1-R_{33}-T_{34})\mu_3-T_{31}\mu_1-T_{32}\mu_2 \right],
\end{gathered}\label{butt4}
\end{equation} 
and
\begin{equation}
\begin{gathered}
I_4=\dfrac{e}{h}\left[ (1-R_{44}-T_{43})\mu_3-T_{41}\mu_1-T_{42}\mu_2 \right].
\end{gathered}\label{butt5}
\end{equation} 
Here, $\mu_{1}$ and $\mu_{2}$ are the chemical potentials of the real leads 1 and 2, respectively. $\mu_3$ is the chemical potential of the virtual lead. Two additional quantities are introduced: $S_f=T_{13}+T_{14}$ and $S_b=T_{23}+T_{24}$. $S_f$ and $S_b$ are the transmission coefficients for an electron appearing from the virtual lead to propagate the non-Hermitian conductor forward in and backward against the direction of the electron flow. There should be no net current in the virtual lead, i.e., $I_3+I_4=0$. Then, we can solve the chemical potentials for the virtual lead $\mu_3=(S_f\mu_1+S_b\mu_2) /(S_f+S_b)$. Then, the net current to the channel 1 is given by 
\begin{equation}
\begin{gathered}
I_1=\dfrac{e}{h}\left[ T_{12}+\dfrac{S_b S_f}{S_b+S_f} \right](\mu_1-\mu_2).
\end{gathered}\label{butt6}
\end{equation}
Accordingly, the conductance is given by $G=(e^2/h)[T_{12}+S_bS_f/(S_b+S_f)]$. From the Landauer formula, 
\begin{equation}
\begin{gathered}
T=\dfrac{G}{(e^2/h)}=T_{12}+\dfrac{S_b S_f}{S_b+S_f}.
\end{gathered}\label{butt7}
\end{equation}
Equation~\eqref{butt7} is valid for a two-terminal configuration no matter how many conducting channels are within the real (virtual) leads. In principle, one must determine $S_b$ and $S_f$ to obtain the transmission $T$, B\"{u}ttiker has discussed some situations in Ref.~\cite{buttiker_prb_1986}. For hinge channels discussed in this work, we have proven that their chirality is preserved in Sec.~\ref{sec4}. Therefore, the backward scattering is strongly suppressed, i.e., $S_b=0$, such that the chemical potential of the virtual lead equals to $\mu_1$. Then, we have $T=T_{12}$ for chiral hinge states. 
\par

\section{Honeycomb laser cavity network}
\label{sec_laser}

The results of 3DSOTIs can be applied to other topological systems like Chern insulators that support chiral boundary states. Stimulated by this, we propose the two-dimensional (2D) laser cavity network on a honeycomb lattice as an easy-to-access platform for experiments. It has been shown that, experimentally~\cite{bandres_sciences_2018} and theoretically~\cite{harari_science_2018}, 2D laser cavity network of a Haldane design processes topologically-protect chiral edge modes winding around the edges. The dynamics of the laser field can be written as
\begin{equation}
\begin{gathered}
i\dfrac{\partial a_{\bm{n}}}{\partial t}=(\omega_0+\omega_{\bm{n}})a_{\bm{n}}\\
+t_1\sum_{\langle \bm{nm}\rangle}b_{\bm{m}}+t_2 e^{i\phi}\sum_{\langle\langle \bm{nm}\rangle\rangle}a_{\bm{m}}-i\gamma a_{\bm{n}}+ i\tilde{g}_a a_{\bm{n}}\\
\quad\\
i\dfrac{\partial b_{\bm{n}}}{\partial t}=(\omega_0+\omega_{\bm{n}})b_{\bm{n}}\\
+t_1\sum_{\langle \bm{nm}\rangle}a_{\bm{m}}+t_2 e^{-i\phi}\sum_{\langle\langle \bm{nm}\rangle\rangle}b_{\bm{m}}-i\gamma b_{\bm{n}}+ i\tilde{g}_b b_{\bm{n}}.
\end{gathered}\label{laser1}
\end{equation}
In Eq.~\eqref{laser1}, $a_{\bm{n}}$ and $b_{\bm{n}}$ are the laser field amplitudes at a site $\bm{n}$ of the A and B sublattices, respectively. $\langle \bm{nm}\rangle$ and $\langle\langle \bm{nm}\rangle\rangle$ denote the nearest-neighbor and next-nearest-neighbor sites, respectively. A resonator is coupled with its nearest-neighbors with a real hopping of constant $t_1$ and its next-nearest-neighbors with a complex hopping of $t_2 e^{i\phi}$. $\phi$ is the Haldane flux parameter. 
\par

Disorders are introduced through the first terms in the right hand side of Eq.~\eqref{laser1}. $\omega_0+\omega_{\bm{n}}$ that is the resonance frequency of resonators at one unit cell $\bm{n}$. Artificially, the resonance frequencies are tunable and depend on the details of the resonators. Here, we model the disorders by setting $\omega_0$ being a constant and $\omega_{\bm{n}}$ distributing randomly and uniformly in the range of $[-W/2,W/2]$ such that $W$ measures the degrees of randomness. $\gamma$ represents the linear loss of a resonator. $\tilde{g}_{a,b}$ are the optical gains can be tuned as well. In Ref.~\cite{cwang_prb_2022}, we have shown that the chiral edge states are Hermitian if
\begin{equation}
\begin{gathered}
-\gamma+(\tilde{g}_a+\tilde{g}_b)/2=0,
\end{gathered}\label{laser2}
\end{equation}
$t_2\neq 0$, and $-\pi<\phi<\pi$. Therefore, if Eq.~\eqref{laser2} is satisfied, we expect the mean transmission coefficient of chiral edge states is equal to 1 even in the presence of disorders. Otherwise, $\langle T\rangle\neq 1$.
\par

\begin{figure}[htbp]
\centering
\includegraphics[width=0.45\textwidth]{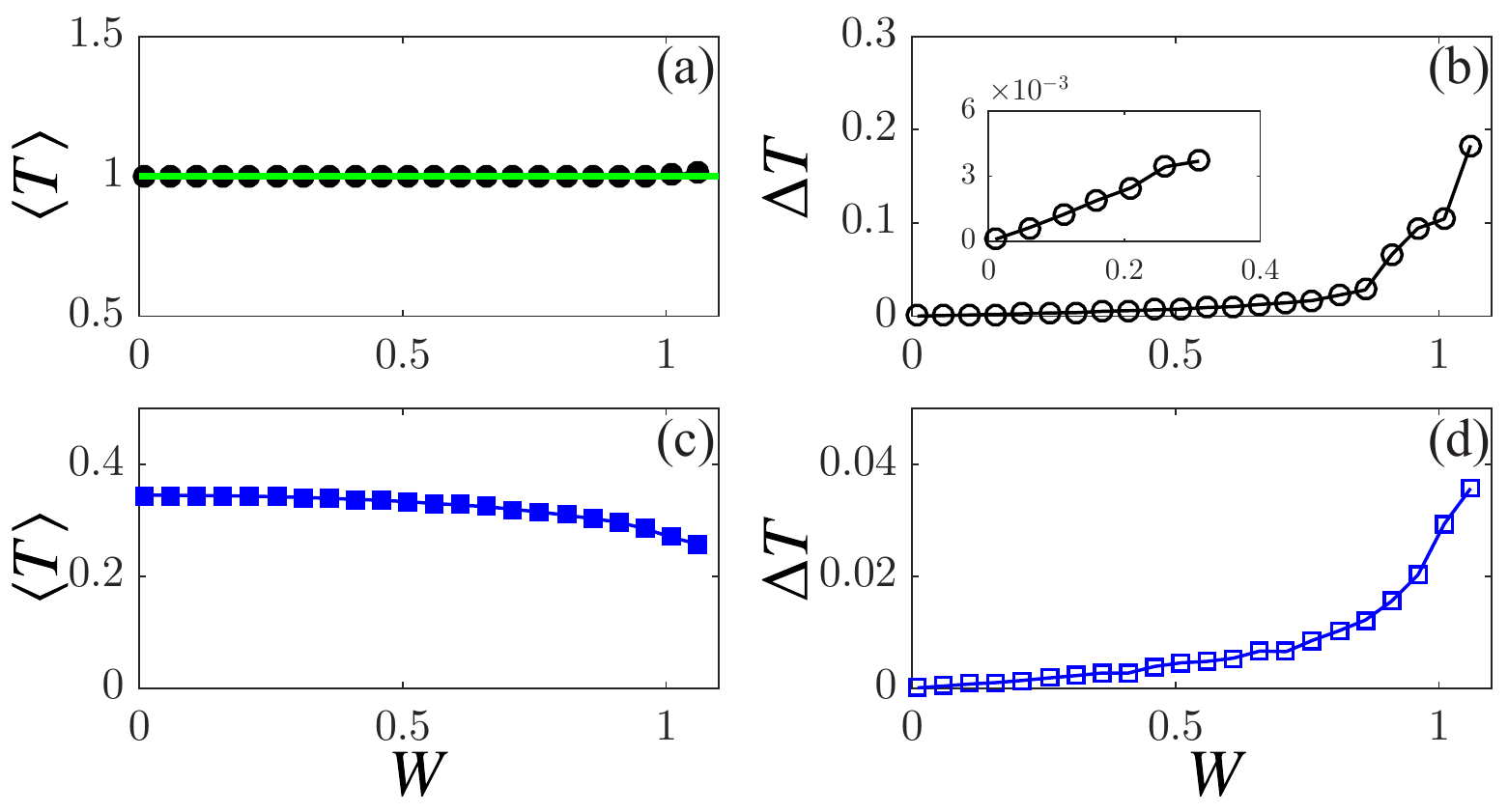}
\caption{(a,b) Mean transmission coefficients $\langle T\rangle$ (the filled symbols) and fluctuations of transmission coefficients $\Delta T$ (the open symbols) as a function of $W$ for the laser cavity network on a honeycomb lattice whose dynamic is governed by Eq.~\eqref{laser1} with $t_1=1$, $t_2=0.1 t_1$, $\phi=\pi/5$, and $\kappa=0.01$. The green solid line is $\langle T\rangle=1$. (c,d) Same as (a,b) but without the optical gain $\tilde{g}_{a,b}=0$ and $\gamma=0.01$.  The length and width of the armchair ribbon are 40 and 30 (in the unit of lattice constant), respectively. Each point here is average over 50 samples.}
\label{fige1}
\end{figure}

To confirm the above prediction, we set $\tilde{g}_a=\gamma+\kappa$ and $\tilde{g}_b=\gamma-\kappa$ with $\kappa$ being a positive number and $\kappa<\gamma$ to construct Hermitian chiral edge states and specifically design the honeycomb lattice to be a two-terminal armchair ribbon with two leads at the two ends along the $x$ direction. We choose $t_1=1$, $t_2=0.1 t_1$, $\phi=\pi/5$, and $\kappa=0.01$, and calculate $\langle T\rangle$ and $\Delta T$ as a function of $W$ for a given frequency $\omega=\omega_0-0.5$. It is shown that the mean transmission coefficient is equals to 1 and the fluctuation of transmission coefficients increases with $W$, see Figs.~\ref{fige1}(a,b). As a comparison, we also display data without the optical gain, i.e., $\tilde{g}_{a,b}=0$, where the chiral edge modes are non-Hermitian and dissipative. In this case, we find $\langle T\rangle<1$, see one representative example in Figs.~\ref{fige1}(c,d). The laser intensity is measurable~\cite{harari_science_2018} such that one can calculate the transmission coefficient $T$ and verify the physics illustrated in this work.   
\par

\end{document}